\documentstyle[12pt]{article}
\begin{document}
\baselineskip=5.0mm
\newcommand{\be} {\begin{equation}}
\newcommand{\ee} {\end{equation}}
\newcommand{\Be} {\begin{eqnarray}}
\newcommand{\Ee} {\end{eqnarray}}
\def\lg{\langle}
\def\rg{\rangle}
\def\a{\alpha}
\def\b{\beta}
\def\g{\gamma}
\def\G{\Gamma}
\def\d{\delta}
\def\D{\Delta}
\def\e{\epsilon}
\def\k{\kappa}
\def\l{\lambda}
\def\L{\Lambda}
\def\om{\omega}
\def\Om{\Omega}
\def\s{\sigma}
\def\t{\tau}
\noindent
\begin{center}
{\large
{\bf
Stochastic models for heterogeneous relaxation: Application to inhomogeneous
optical lineshapes
}}
\vspace{0.5cm}

\noindent
{\bf Gregor Diezemann, Gerald Hinze and Hans Sillescu} \\
\vspace{0.5cm}

\noindent
Institut f\"ur Physikalische Chemie, Universit\"at Mainz,
Welderweg 15, 55099 Mainz, FRG
\\
\end{center}
\vspace{0.5cm}
\noindent
{\it
Dynamic heterogeneity has often been modeled by assuming that a single-particle
observable, fluctuating at a molecular scale, is influenced by its
coupling to environmental variables fluctuating on a second, perhaps slower,
time scale.
Starting from the most simple Gaussian Markov process we model the exchange
between 'slow' and 'fast' environments by treating the fluctuating
single-particle variable as a projection from a higher-dimensional Markov
process.
The moments of the resulting stochastic process are calculated from the
corresponding Master equations or Langevin equations, depending on the model.
The calculations show the importance of the way to treat exchange processes.
The resulting stochastic process is non-Markovian for all models.
However, the deviations from a Gaussian behavior depend on the details of
the models.
A comparison of our results with other model treatments and experiments
should provide further insight into the concept of dynamic heterogeneity.
}

\vspace{1.0cm}
\subsection*{I. Introduction}
In spectroscopy inhomogeneous lineshapes are observed if individual
molecules within a sample give rise to different spectral components.
This can be observed even in isotropic systems if the relevant coupling
depends on molecular variables, like orientations, that differ from each
other.
Another, more important source for inhomogeneous line broadening is provided
by intermolecular couplings with the environment.
These give rise to dynamic fluctuations and can thus be explored by lineshape
studies.
Classical examples are given by the motional or exchange narrowing of
NMR spectra\cite{abra}. The Anderson-Weiss theory\cite{AW53} of narrowing
Gaussian lineshapes has been a model for many later treatments
although it applies to homogeneous systems only.
Furthermore, typical 'slow motion' lineshapes in NMR are far from being
Gaussian.
However, for our purposes, the Gaussian approximation provides a good
starting point since it implies the well-known Ornstein-Uhlenbeck (OU)
process\cite{vkamp}.
This allows one to introduce non-Gaussian perturbations in a transparent way.

A shortcoming of the OU process is that the two-time correlation function
(2t-CF) of this Gaussian Markov process decays exponentially\cite{vkamp}.
In complex systems like supercooled liquids, however, typically a
non-exponential decay of 2t-CFs is observed.
This forces us to consider non-Markovian models for the molecular observables.
There are other fields in chemical physics where, starting from Gaussian
fluctuations, non-Markovian effects have been incorporated in order to account
for a more complex dynamics.
Prominent examples are the problem of escape through a fluctuating
bottleneck\cite{zwanzig92,WW93,BS98}, first order chemical
reactions\cite{AH83,WW94} and single molecule spectroscopy\cite{WW95}.

In our treatment, we closely follow the theory of solvation dynamics in
time-resolved optical spectroscopy\cite{LYM87,S397} that has also previously
been applied to treat heterogeneity in terms of observed deviations from
Markovian and Gaussian behavior\cite{ranko01a,ranko01b,dieze01}.
It should be noted that the average frequency, $\lg\om(t)\rg$, remains
constant in essentially all dynamical NMR and ESR spectra but it changes due to
time-dependent Stokes shifts in time-resolved optical spectra.
This can result in time-dependent optical linewidths that have been related
to dynamic heterogeneity\cite{ranko01a,WR00}.

In the following Section, we briefly review the theory of time-dependent
optical lineshapes in the form that will be used in the calculations.
In Section III we treat the optical transition frequency $\om(t)$ as a
non-Markovian stochastic process, defined as the projection from a composite
Markov process. For simplicity, we allow for two values of an additional
environmental variable, $\e(t)$, to treat exchange between a 'slow' ($\e_s$)
and a 'fast' ($\e_f$) environment.
Exchange in two (or multiple) state models has been treated a long time ago
in relation with rotational\cite{AU67,BP65,sill71,sill96} and
translational\cite{kaerger69,CS97} diffusion in complex systems.
Here, we concentrate on the question regarding deviations from both,
Gaussian and Markovian, behavior.
The paper closes with a discussion of our results and conclusions in
Section IV.
\subsection*{II. Short survey of the theory of optical lineshapes}
This section consists of a brief review of the theory of time-dependent
optical lineshapes in the inhomogeneous limit.
In case of triplet state solvation all effects related to internal
conversion, intersystem crossing, and the following electronic relaxation are
assumed to be much faster than the dynamical features monitored experimentally.
We mainly follow the general theoretical framework of
refs.\cite{LYM87,S397,dieze01} in the following.
There, it has been shown that the time-dependent signal for a single
chromophore immersed in a bath of solvent molecules can be written as:
\be\label{It.def}
I(\om,t)= \int_{-\infty}^\infty d\om_0P(\om,t|\om_0)p_0(\om_0)
\ee
Here, the transition frequencies $\om(t)$ depend on all relative distances
and orientations between the chromophore and the solvent molecules, which in
turn depend on time.
In Eq.(\ref{It.def}) $P(\om,t|\om_0)$ denotes the conditional probability of
finding a transition frequency $\om$ at time $t$, given it was $\om_0$
at time $t\!=\!0$.
The initial probability $p_0(\om)$ is given by the emission spectrum at time
$t\!=\!0$, $I(\om,0)$, because of $P(\om,0|\om_0)\!=\!\d(\om-\om_0)$.
Thus, observation of $I(\om,t)$, Eq.(\ref{It.def}), as a function of time is
interpreted as following the trace of the initial non-equilibrium emission
spectrum towards the equilibrium steady-state emission spectrum
$I(\om,\infty)=p^{eq}(\om)=P(\om,\infty|\om_0)$.
Furthermore, it is assumed that the time-independent spectra are well
approximated by Gaussians,
\be\label{p0.eq}
p_0(\om)=\frac{1}{\sqrt{2\pi\s_0^2}}e^{-(\om-\D)^2/(2\s_0^2)}
\quad\mbox{and}\quad
p^{eq}(\om)=\frac{1}{\sqrt{2\pi\s_\infty^2}}e^{-\om^2/(2\s_\infty^2)}
\ee
where $\D=\lg\om(0)\rg$, $\s_0^2=\lg(\om(0)-\D)^2\rg$ and
$\s_\infty^2=\lg\om(\infty)^2\rg$. The mean value $\lg\om(\infty)\rg$ has
been set to zero without loss of generality.
Therefore, the overall red-shift is given by $\D$.
A Gaussian approximation for the conditional probability,
\be\label{Pt.gauss}
P(\om,t|\om_0)=\frac{1}{\sqrt{2\pi\s_\infty^2(1-C(t)^2)}}
    \exp{\left(-\frac{[\om-\om_0C(t)]^2}
		     {2\s_\infty^2(1-C(t)^2)}\right)}
\ee
follows from a second order cumulant expansion with the Stokes shift
correlation function given by $C(t)=\lg\om(t)\om(0)\rg/\s_\infty^2$.
Using Eq.(\ref{Pt.gauss}) in Eq.(\ref{It.def}) for the time-dependent
spectrum again gives a Gaussian for $I(\om,t)$ with a time-dependent mean and
a second moment
\be\label{moms.hom}
\lg\om(t)\rg = \D\: C(t)
\quad\mbox{and}\quad
\lg\om^2(t)\rg = \s_\infty^2+(\D^2+\s_0^2-\s_\infty^2)C(t)^2
\ee
yielding a time-dependent variance
\be\label{sig.t.hom}
\s^2(t) = \lg\om^2(t)\rg - \lg\om(t)\rg^2
	 = \s_\infty^2+(\s_0^2-\s_\infty^2)C(t)^2
\ee
This variance shows a monotonic behavior as a function of time in that it
changes monotonously from $\s^2(0)=\s_0^2$ to $\s^2(\infty)=\s_\infty^2$.

It is precisely the deviations from the simple relation (\ref{sig.t.hom})
observed in the triplet state solvation studies on supercooled
liquids\cite{rankoreview,WR00} that has led to the interpretations of
the results in terms of heterogeneous dynamics.
In supercooled liquids, the Stokes shift correlation function decays
non-exponentially and may be written as
\be\label{Ct.lambda}
C(t)=\int\!d\l p(\l) e^{-\l t}
\ee
with relaxation rates $\l$ and a distribution function $p(\l)$.
Performing a so-called inhomogeneous cumulant expansion of the conditional
probability, one finds\cite{dieze01}
\Be\label{P.ic}
&&P(\om,t|\om_0)=\int\!d\l p(\l)P_{(\l)}(\om,t|\om_0)\nonumber\\
&&P_{(\l)}(\om,t|\om_0)=
\frac{1}{\sqrt{2\pi\s_\infty^2(1-e^{-2\l t})}}
    \exp{\left(-\frac{[\om-\om_0e^{-\l t}]^2}
		     {2\s_\infty^2(1-e^{-2\l t})}\right)}
\Ee
instead of Eq.(\ref{Pt.gauss}).
For the first moment, one again finds $\lg\om(t)\rg = \D\: C(t)$,
cf. Eq.(\ref{moms.hom}), whereas the variance is given by
\be\label{sig.t.het}
\s^2(t)=\s_\infty^2+(\s_0^2-\s_\infty^2)C(t)^2
	 +\D^2\left[C(2t)-C(t)^2\right]
\ee
This expression was found to give excellent agreement with the experimentally
observed linewidth\cite{ranko01a}.

It is seen that according to Eq.(\ref{P.ic}) the stochastic process
$\om(t)$ can be viewed as originating from a superposition of Gaussian
processes.
This means that $\om(t)$ is modeled as a {\it non-Markovian Gaussian}
stochastic process.
In the next Section we discuss various exchange models, with
particular emphasis on the deviations from Gaussian behavior.
\subsection*{III. Calculation of frequency moments in exchange models}
The simplest stochastic model for a process like solvation is a Gaussian
Markov process, i.e. the OU-process.
If the transition frequency $\om(t)$ is modeled as an OU-process, one may
consider the Langevin equation (LE):
\be\label{LE.OU}
{\dot \om}(t)=-\g\om(t) + \xi(t)
\ee
where $\xi(t)$ is a delta-correlated white noise,
$\lg\xi(t)\rg\!=\!0$, $\lg\xi(t)\xi(t')\rg\!=\!2\s_\infty^2\g\d(t-t')$ and
$\g$ denotes the damping rate.
The LE, Eq.(\ref{LE.OU}), approximates the spectral diffusion process as
occuring on harmonic potential-energy surfaces.
The transition frequency spreads in a restoring potential $V\!=\!(\g/2)\om^2$.
Alternatively, one can consider the corresponding Fokker-Planck equation
(FPE) for the conditional probability:
\be\label{FPE.OU}
{\dot P}(\om,t|\om_0) =
{\hat \Pi}^{FP}(\om)P(\om,t|\om_0)
\quad\mbox{with}\quad
{\hat \Pi}^{FP}(\om)=\g\left[\frac{\partial}{\partial \om}\om +
		     \s_\infty^2\frac{\partial^2}{\partial \om^2}\right]
\ee
The resulting expression for $P(\om,t|\om_0)$ is given by
Eq.(\ref{Pt.gauss}) with $C(t)\!=\!\exp{(-\g t)}$. Therefore, in order to
have a model which yields a Stokes shift correlation function decaying
non-exponentially according to Eq.(\ref{Ct.lambda}), one has to consider
$\om(t)$ as a non-Markovian stochastic process.
In the present paper, this is achieved by considering exchange models.
As already indicated in the introduction, we restrict ourselves to two-state
models, because these allow for analytic solutions.
The idea underlying such models can most easily be summarized in the following
way. We consider the solute to experience various surroundings characterized
by an environmental variable $\e_i$, $i\in\{s,f\}$.
Thus, in a given environment, the spectral diffusion of the optical transition
frequencies $\om(t)$ occurs with a decay constant $\g_i\!=\!\g(\e_i)$.
However, the environments are not viewed to be static, but rather fluctuate in
the course of time. These environmental fluctuations are characterized by
exchange rates $\k_{i,k}\!=\!\k(\e_i|\e_k)$. In order to further simplify
the calculation, we take the exchange rates to be equal,
$\k_{s,f}\!=\!\k_{f,s}\!=\!\g_x$, implying $p^{eq}_i\!=\!p^{eq}(\e_i)\!=\!1/2$.
The generalization to a larger number of states is trivial. Also other models, like energy
landscape models\cite{dieze97}, can be considered\cite{dieze01}.

Even though we restrict ourselves to simple two-state models, we have to face
the ambiguity concerning the way of {\it how} to incorporate the exchange
between the fast and slow environments into the model.
In the following, we consider two different, but in some way strongly related
ways of treating the rate exchange.
\subsubsection*{A. Master equation models}
One class of models allowing to treat rate exchange processes can be defined
by supplementing the FPEs in the slow and fast environments with some terms
representing the exchange process.
This way one obtains rate equations for the conditional probability
$P_{i,k}(\om,t|\om_0)\!\equiv\!P(\om,\e_i,t|\om_0,\e_k)$ for finding the
pair $\{\om,\e_i\}$ at time $t$, given it had values $\{\om_0,\e_k\}$ at
$t\!=\!0$.
Still, we have to make some additional asumptions about what happens
with the value of $\om$ during the exchange process.
Two limiting choices for this are the following\cite{BP65,sill96}.
Either one assumes that $\om$ does not change at all during a $\e_i\to\e_k$
transition.
Alternatively, one allows $\om$ to randomize completely during such a
transition, meaning that after the transition $\om$ can take any value
according to its statistical weight $p^{eq}(\om)$.
Let us start with the first choice.
This is the same assumption that has often been used in context of models
for rotational and translational motions in supercooled
liquids\cite{sill96,CS97} and also in treatments of spectral
diffusion\cite{ranko01b}.

We abbreviate the FP-operators in the fast and in the slow environments by
${\hat \Pi}^{FP}_i(\om)$. These operators are defined in Eq.(\ref{FPE.OU}),
but with $\g_i$ replacing $\g$. We have to solve the coupled FPEs:
\be\label{ME.sf}
{\dot P}_{i,l}(\om,t|\om_0) =
 {\hat \Pi}_{i}^{FP}(\om)P_{i,l}(\om,t|\om_0)
 - \g_x\left[P_{i,l}(\om,t|\om_0) - P_{k,l}(\om,t|\om_0)\right]
\quad;\quad i\neq k
\ee
Due to the coupling of the FPEs, the resulting $P_{i,l}(\om,t|\om_0)$ are
not Gaussians. Furthermore, an analytical solution of Eq.(\ref{ME.sf}) is
not feasible.
In order to calculate the spectral moments or any other observable, an
expansion in terms of the eigenfunctions of the FP operator\cite{risken} is
utilized,
$P_{i,k}(\om,t|\om_0)\!=\!e^{-\bar{\om}^2}\sum_{n=0}^\infty
N_n H_n(\bar{\om})H_n(\bar{\om}_0)G^{(n)}_{i,k}(t)$,
where $H_n(z)$ are Hermite polynomials, $\bar{\om}^2=\om^2/2\s_\infty^2$
and $N_n^{-1}\!=\!2^nn!(2\pi\s_\infty^2)^{1/2}$.
Using this expression in the FPEs, Eq.(\ref{ME.sf}), results in the following
equations for the Green's functions:
\be\label{ME.Gn}
\frac{\partial}{\partial t} G^{(n)}_{i,l}(t) =
-\left(n\g_i\right) G^{(n)}_{i,l}(t)
-\g_x\left[G^{(n)}_{i,l}(t) - G^{(n)}_{k,l}(t)\right]
\quad\mbox{where}\quad k\neq i
\ee
which can be solved easily.
In the following, we assume $\s_0=\s_\infty$ for simplicity.
For the lowest-order spectral moments one finds, after performing the
$\om$-integrations:
\Be\label{m.1.2}
&&\lg\om(t)\rg=\D\: C_1(t)\nonumber\\
&&\lg\om^2(t)\rg=\s_\infty^2+\D^2 C_2(t)
\Ee
Furthermore, one has $\lg\om^3(t)\rg=3\D\:\s_\infty^2C_1(t)+\D^3 C_3(t)$ and
$\lg\om^4(t)\rg=3\s_\infty^4+6\s_\infty^2\D^2C_2(t)+\D^4C_4(t)$.
In principle, all moments are accessible this way, although the calculations
become rather tedious. It is important to notice at this point that the
$n^{\rm th}$ moment $\lg\om^n(t)\rg$ is determined by the functions
\be\label{Cn.t.def}
C_n(t)=\frac{1}{2}\sum_{i,k}G^{(n)}_{i,k}(t)=
p^{(n)}_s\exp{[-\g^{(n)}_st]}+p^{(n)}_f\exp{[-\g^{(n)}_ft]}
\ee
Here, the $p^{(n)}_i$ and $\g^{(n)}_i$ are determined by the solution of
Eq.(\ref{ME.Gn}):
\Be\label{ME.eigen}
&&p^{(n)}_{s/f}=\frac{1}{2}\left( 1\pm \frac{2\g_x}{Z_n}\right)\quad
\g^{(n)}_{s/f}=\frac{1}{2}\left(2\g_x+n(\g_f+\g_s) \mp Z_n\right)\\
&&\mbox{with}\quad
Z_n=\sqrt{\left(n^2(\g_f-\g_s)^2+4\g_x^2 \right)}\nonumber
\Ee

In Fig.1a we plotted the Stokes shift correlation function
$C_1(t)\!\equiv\!C(t)$ versus time. For the parameters chosen it is clearly
seen that mainly the long time decay is affected by a finite exchange rate
$\g_x$.
Fig.1b shows the time-dependent part of the variance,
$\s_{sc.}^2(t)\!=\![\s^2(t)-\s_\infty^2]/\D^2$ versus time for the same
parameters.
Additionally shown is the result of the Gaussian approximation,
$C_1(2t)-C_1(t)^2$, cf. Eq.(\ref{sig.t.het}). Remember, that this expression
is in excellent agreement with the available experimental data\cite{WR00}.
It is clearly seen that $\s_{sc.}^2(t)$ exceeds this expression, in particular
in the time range, where both functions exhibit their maximum.
It is exactly this excess observed for this particular model that previously
led to the incorrect conclusion that the experimental data are only
compatible with a long-lived dynamic heterogeneity\cite{ranko01b}.
In the present model the differences between $\s_{sc.}^2(t)$ and
$C_1(2t)-C_1(t)^2$ originate {\it solely} from the fact, that the
stochastic process $\om(t)$ is not only non-Markovian, but also
non-Gaussian.

From the expressions given in Eq.(\ref{ME.eigen}) for $n\!\gg\!1$ one
approximately has $\g^{(n)}_{s/f}\!\simeq\!\g_x+n\g_{s/f}$ and
$p^{(n)}_{s/f}\!\simeq\!1/2\pm \g_x/(n(\g_f-\g_s))$.
Therefore, using the parameters of Fig.1, one finds that even for $n\!=\!100$
the exchange rate $\g_x$ still contributes on the order of $30\%$ to
$\g^{(100)}_s$.
In Fig.2 we show the quantities $\zeta_n(t)=C_n(t)-C_1(nt)$. These give
a measure for the deviations from a Gaussian behavior because in that case
one has $\zeta_n(t)\equiv 0$ for all $n\!>\!0$.
This plot nicely illustrates the fact just discussed.
With increasing $n$ the maximum $\zeta_n(t)$ is shifted towards shorter times
$t_{max,n}$, Fig.2a.
However, one can see that the maximum of $\zeta_n(t)$ roughly occurs at
$n\!\times\!t_{max,n}\!\sim\!\g_x^{-1}$, as shown in Fig.2b.
This shows that not only the first few moments are affected by the deviations
from Gaussian behavior in this model, but that these deviations even increase
with $n$.

In the above treatment we considered the case in which the transition
frequencies $\om(t)$ do not change during a $\e_i\to\e_k$ transition.
We already discussed that another extreme scenario would be defined by
allowing $\om(t)$ to randomize during such a transition.
We just mention that for this model the eigenvalues for $n\!=\!0$ remain the
same as in the case considered above.
For finite $n$, however, one has $p^{(n)}_{s/f}=1/2$ and
$\g^{(n)}_{s/f}=\g_x+n\g_{s/f}$.
Thus, the qualitative features are not very different in this case.
In particular, in these exchange models the stochastic process $\om(t)$ is
non-Markovian {\it and} non-Gaussian.
\subsubsection*{B. Langevin equation models}
Instead of starting from the coupled FPEs, Eq.(\ref{ME.sf}), one can also
define an exchange model by considering coupled Langevin equations.
In such models one associates a value $\om_i(t)$ with either of the states
$i\!=\!s,f$ and introduces an additional exchange term, cf. Eq.(\ref{LE.OU}):
\be\label{LE.sf}
{\dot \om}_i(t)=-\left(\g_i+\g_x\right)\om_i(t)
				    +\g_x\om_k(t) + \xi_i(t)
\ee
where again $i\!\neq\!k$ and $\xi_i(t)$ is a delta-correlated white noise,
meaning that $\lg\xi_i(t)\rg\!=\!0$ and
$\lg\xi_i(t)\xi_k(t')\rg\!=\!\G_{i,k}\d(t-t')$.
Note, that here we do not have different choices for the incorporation of
exchange effects, since by definition the LE model involves diffusive motions
only.
Therefore, in such a model $\om$ changes by an infinitesimal amount during an
$\e_i\!\to\!\e_k$ transition.
The solution of Eq.(\ref{LE.sf}) results in two-dimensional Gaussians for
the conditional probabilities.
In order to show how the results are related to the ones obtained for the FPEs,
we proceed in the following way.
We introduce the eigenvalues $\l_\a$ and the corresponding eigenvectors
$|\a\rg$ of the matrix ${\bf W}$ defined in Eq.(\ref{LE.sf}).
In such a notation we have $W_{i,k}=\lg i|{\bf W}|k\rg$,
$W_{s,s}=-\g_s-\g_x$, $W_{f,f}=-\g_f-\g_x$, $W_{s,f}=W_{f,s}=\g_x$
and $\om_i(t)=\lg i|\om(t)\rg$. Correspondingly, using
$\sum_i|i\rg\lg i|=\sum_\a|\a\rg\lg\a|=1$ and the orthonormality of the
vectors, we have $\om_i(t)=\sum_\a\lg i|\a\rg\om_\a(t)$
and $W_{i,k}=\sum_\a\lg i|\a\rg\l_\a\lg\a|k\rg$.
This yields the decoupled LEs:
\be\label{LE.lambda}
{\dot \om}_\a(t)=\l_\a\om_\a(t) + \xi_\a(t)
\ee
from which it is imediately evident that the solutions for the corresponding
$P_\a(\om,t|\om_0)$\cite{vkamp} are Gaussians determined by the correlation
functions $C_\a(t)\!=\!\lg\om_\a(t)\om_\a(0)\rg$.
In Eq.(\ref{LE.lambda}), the projections of the noise give
$\lg\xi_\a(t)\xi_\a(t')\rg\!=\!\G_\a\d(t-t')$.
Choosing $\G_{i,k}=-2\s_\infty W_{i,k}$ yields $\G_\a=-2\l_\a$.
For the moments one finds:
\be\label{Moms.lambda}
\lg\om_\a(t)\rg=\lg\om_\a(0)\rg e^{\l_\a t}
\quad\mbox{and}\quad
\lg\om_\a^2(t)\rg=\lg\om_\a^2(0)\rg e^{2\l_\a t}
+\s_\infty^2\left(1-e^{2\l_\a t}\right)
\ee
and the correlation function is given by
$C_\a(t)\!=\!\lg\om_\a^2(0)\rg e^{\l_\a t}$.
For an arbitrary observable the expectation value is given by
$\lg A(t)\rg=(1/2)\sum_\a\lg A_\a(t)\rg$, and thus
\be\label{Moms.res}
\lg\om(t)\rg=\D\: C_1(t)
\quad\mbox{and}\quad
\lg\om^2(t)\rg=\s_\infty^2+\D^2C_1(2t)
\ee
where the $C_n(t)$ are defined in Eq.(\ref{Cn.t.def}).
Therefore, the time-dependent variance in this model is given by
$\s^2(t)\!=\!\s^2_\infty+\D^2[C_1(2t)-C_1(t)^2]$, cf. Eq.(\ref{sig.t.het}).
In Fig.1b, $C_1(2t)-C_1(t)^2$ is shown as the dashed line.
The mathematical reason for the behavior of the moments, Eq.(\ref{Moms.res}),
lies in the fact that the matrix ${\bf W}$ defined in Eq.(\ref{LE.sf}) is
identical to the one for $G^{(1)}_{i,k}(t)$ defined in Eq.(\ref{ME.Gn}).
This means that the eigenvalues $\l_a$ are just given by
$\l_{s/f}\!=\!-\g^{(1)}_{s/f}$ according to Eq.(\ref{ME.eigen}).
The conclusion from this observation is that the Langevin model considered
here is nothing but the Gaussian version of the master equation model
studied above.
A simple calculation shows that the LEs, Eq.(\ref{LE.sf}), considered here
corresponds to the coupled FPEs ${\dot P}_{i,l}(\om,t|\om_0)\!=
\!\sum_k{\hat \Pi}_{ik}^{FP}(\om)P_{k,l}(\om,t|\om_0)$.
Here, ${\hat \Pi}_{ik}^{FP}(\om)$ is obtained from Eq.(\ref{FPE.OU}) by the
replacement $\g\!\to\!(-W_{i,k})$.
The important point is that in this model $\om(t)$ is a
{\it non-Markovian Gaussian} process.
\subsection*{IV. Discussion and Conclusions}
In the preceeding section we have analyzed simple exchange models that
allow one to account for fluctuations in dynamic heterogeneous systems.
In the particular example of solvation dynamics we have focussed on the
calculation of the lowest-order moments of the transition frequency
distributions.
Despite its simplicity this example allows to discuss various aspects of
dynamic heterogeneity.
In particular, without exchange the transition frequency $\om(t)$ constitutes
a Gaussian Markov process, the OU process. This is among the simplest
stochastic processes one can study\cite{vkamp}.

We have investigated different variants of the models to incorporate the way
in which exchange processes change the temporal evolution of the conditional
probability. These exchange processes are modeled as transitions between
different environments, characterized by different values of an additional
stochastic process $\e(t)$.
In the master equation models one has to make specific assumptions about the
effect of exchange on the stochastic process $\om(t)$.
We discussed only two extreme choices. One choice, which also has
been used in many earlier investigations, corresponds to the assumption that
the processes $\om(t)$ and $\e(t)$ are uncorrelated.
This means that $\om(t)$ is not at all affected by a change in $\e(t)$, i.e.
by exchange.
We have shown that the equations determining the moments $\lg\om^n(t)\rg$
can then be solved analytically. This allows to discuss their behavior in some
detail.
A very important finding is that in this case the process $\om(t)$ is
{\it neither Gaussian nor Markovian} for finite exchange rates $\g_x$.
In contrast, when studying a Langevin model, we found that in that case
$\om(t)$ also is {\it non-Markovian  but Gaussian}.
Therefore, using this model yields excellent agreement with the experimental
data\cite{WR00} for arbitrary choices of the exchange rates.
This, however, is not astonishing because according to Eq.(\ref{P.ic})
{\it any} model that yields a description of $\om(t)$ in terms of a
non-Markovian Gaussian stochastic process will give results compatible with
the experimental data.

The most important point in the discussion of the lifetime of the dynamic
heterogeneities appears to be the question as to how to quantify this
lifetime. It has been stressed earlier that this cannot be achieved by any
two-time correlation function\cite{vigo98}.
This fact stimulated the development of NMR techniques to study higher-order
time correlation functions\cite{SRS91}.
A comparison of the conditional probabilities $P(\om,t|\om_0)$ with
two-dimensional NMR spectra\cite{SRS94} shows that these are identically the
same if $\om$ is reinterpreted as the spin-precession frequency.
The mentioned time dependence of $\lg\om(t)\rg$ that is absent in NMR
does not enter the discussion here, because this time dependence does not
alter $P(\om,t|\om_0)$. Thus, even if the full conditional probability would
be known, one could not obtain information about the heterogeneity lifetime.
However, knowledge of higher moments would allow to better specify the
detailed 'geometric' properties of the spectral diffusion process.
In particular, one could get information about possible deviations from
diffusive behavior, which was our starting point in this paper.
It is exactly this issue which can be treated very efficiently by 2d-NMR
methods. One can clearly distinguish between rotational diffusion and
rotational jump models\cite{SRS94}.

The comparison between the spectral diffusion process and the reorientational
motion of molecules in supercooled liquids further shows the different
impact of exchange on various models.
Rotational diffusion (or also rotations by finite jumps) represents a Markov
process. This process, however, is not Gaussian.
Therefore, in this example it hardly matters in which way exchange processes
are incorporated. By contrast, in the example of spectral diffusion, different
ways of treating exchange give different results concerning the properties of
the resulting stochastic process. As discussed above, using master equation
models usually will destroy the Gaussian properties of $\om(t)$.
Therefore, in this case one has a larger variety of modifications that can be
achieved by considering various exchange models.

In conclusion, we have investigated different scenarios for exchange processes
in dynamically heterogeneous systems.
Starting from a Gaussian Markov process, we have shown that different
assumptions concerning the coupling of $\om(t)$ to the exchange processes
yield different results. In particular, if the values of $\om$ change
continuously during an exchange process the resulting stochastic process
is non-Markovian but still Gaussian. Such a 'diffusive' change of $\om(t)$
during exchange of course means that the effect of 'exchange' ($\g_x$) is very
similar to the one of 'relaxation' ($\g_{s/f}$).
In contrast, if exchange does not at all affect the values of $\om$ or if it
causes them to change discontinuously also the Gaussian properties of the
resulting stochastic process is lost.
In any case, deviations from a Gaussian behavior are not good candidates
for the determination of heterogeneity lifetimes. Still, they may be helpful
in discriminating among various variants of exchange models.
\subsubsection*{Acknowledgement}
We are grateful to R. B\"ohmer for fruitful discussions and to the Deutsche
Forschungsgemeinschaft for continuing support of our research on dynamic
heterogeneities via the Sonderforschungsbereich 262.

\section*{Figure captions}
\begin{description}
\item[Fig.1 : ] The first and second moments versus time for the
master equation model.\\
{\bf a}: The Stokes shift correlation function $C_1(t)$ according to
Eq.(\ref{Cn.t.def}) versus time for $\g_s\!=\!0.1s^{-1}$,
$\g_f\!=\!10.0s^{-1}$, $\g_x\!=\!0.33s^{-1}$ (full line) and
$\g_x\!=\!0$ (dotted line).\\
{\bf b}: The time-dependent part of variance,
$\s^2_{sc.}(t)\!=\![\s^2(t)-\s_\infty^2]/\D^2$, versus time (full line) for
$\g_x\!=\!0.33s^{-1}$. Also shown is $C_1(2t)-C_1(t)^2$ as dashed line.
This corresponds to the time-dependent part of the variance in the Langevin
equation model.
The $\g_s$ and $\g_f$ are the same as in (a).
\item[Fig.2 : ] The quantities $\zeta_n(t)\!=\!C_n(t)-C_1(nt)$ versus time for
the same parameters as in Fig.1 ($\g_x\!=\!0.33s^{-1}$).
{\bf a}: $\zeta_2(t)$: full line, $\zeta_{10}(t)$: dashed line,
$\zeta_{20}(t)$: dot-dashed line and $\zeta_{100}(t)$: dotted line.\\
{\bf b}: The same $\zeta_n(t)$ as in (a) are plotted versus the scaled time
$(nt)$. The arrow indicates $\g_x^{-1}\!=\!3.0s$.
\end{description}

\newpage
\begin{thebibliography}{XX}
\bibitem{abra}
A. Abragam; {\it The Principles of Nuclar Magnetism}, Clarendon, Oxford (1961)
\bibitem{AW53}
P.W. Anderson and P.R. Weiss; Rev. Mod. Phys. {\bf 25} 269 (1953)
\bibitem{vkamp}
N.G. van Kampen: {\it Stochastic Processes in Physics and Chemistry},
North-Holland, Amsterdam, New York, Oxford (1981)
\bibitem{zwanzig92}
R. Zwanzig; J. Chem. Phys. {\bf 97} 3587 (1992)
\bibitem{WW93}
J. Wang and P.G. Wolynes; Chem. Phys. Lett. {\bf 212} 427 (1993)
\bibitem{BS98}
D.J. Bicout and A. Szabo; J. Chem. Phys. {\bf 108} 5491 (1998)
\bibitem{AH83}
N. Agmon and J.J. Hopfield; J. Chem. Phys. {\bf 78} 6947 (1983)
\bibitem{WW94}
J. Wang and P.G. Wolynes; Chem. Phys. {\bf 180} 141 (1994)
\bibitem{WW95}
J. Wang and P.G. Wolynes; Phys. Rev. Lett. {\bf 74} 4317 (1995)
\bibitem{LYM87}
R.G. Loring, Y.J. Yan and S. Mukamel; J. Chem. Phys. {\bf 87} 5840 (1987)
\bibitem{S397}
M.D. Stephens, J.G. Saven und J.L. Skinner; J. Chem. Phys. {\bf 106} 2129
(1997)
\bibitem{ranko01a}
R. Richert; J. Chem. Phys. {\bf 114} 7471 (2001)
\bibitem{ranko01b}
R. Richert; J. Chem. Phys. {\bf 115} 1429 (2001)
\bibitem{dieze01}
G. Diezemann; manuscript submitted to J. Chem. Phys.
\bibitem{WR00}
H.Wendt and R.Richert; Phys. Rev. {\bf E 61} 1722 (2000)
\bibitem{AU67}
J.E. Anderson and R. Ullman; J. Chem. Phys. {\bf 47} 2178 (1967)
\bibitem{BP65}
D. Becker and H. Pfeifer; Ann. Phys. {\bf 16} 22 (1965)
\bibitem{sill71}
H. Sillescu; J. Chem. Phys. {\bf 54} 2110 (1971)
\bibitem{sill96}
H. Sillescu; J. Chem. Phys. {\bf 104} 4877 (1996)
\bibitem{kaerger69}
J. K\"arger; Ann. Physik, 7. Folge {\bf 24} 1 (1969)
\bibitem{CS97}
I. Chang and H. Sillescu; J. Phys. Chem. {\bf 101} 8794 (1997);
E. Bartsch, T. Jahr, A. Veniaminov and H. Sillescu;
J. Phys. IV, France {\bf 10} 289 (2000)
\bibitem{rankoreview}
R. Richert; J. Chem. Phys. {\bf 113} 8404 (2000)
\bibitem{dieze97}
G. Diezemann; J. Chem. Phys. {\bf 107} 10112 (1997)
\bibitem{risken}
H. Risken: {\it The Fokker-Planck Equation}, Springer, Berlin (1984)
\bibitem{vigo98}
R. B\"ohmer, R.V. Chamberlin, G. Diezemann, B.Geil, A. Heuer, G. Hinze,
S.C. Kuebler, R. Richert, B. Schiener, H. Sillescu, H.W. Spiess and
M. Wilhelm; J. Non-Cryst. Solids {\bf 235-237} 1 (1998)
\bibitem{SRS91}
K. Schmidt-Rohr and H.W. Spiess; Phys. Rev. Lett. {\bf 66} 3020 (1991)
\bibitem{SRS94}
K. Schmidt-Rohr and H.W. Spiess; {\it Multidimensional Solid-State NMR},
Academic Press, London 1994
\end{thebibliography}
\end{document}